\documentclass{aa}    

\def\simless{\mathbin{\lower 3pt\hbox
     {$\rlap{\raise 5pt\hbox{$\char'074$}}\mathchar"7218$}}} 
\def\simgreat{\mathbin{\lower 3pt\hbox
     {$\rlap{\raise 5pt\hbox{$\char'076$}}\mathchar"7218$}}} 

\usepackage {psfig, epsfig}

\begin{document}


\title{Spitzer IRS spectroscopy of 3CR radio galaxies and quasars:
Testing the unified schemes}

\author {Martin~Haas\inst{1}
  \and Ralf~Siebenmorgen\inst{2}
  \and Bernhard~Schulz\inst{3}
  \and Endrik~Kr\"ugel\inst{4}
  \and Rolf~Chini\inst{1}
}

\institute{Astronomisches Institut, Ruhr-Universit\"at, 
        Universit\"atsstr. 150\,/\,NA7, D-44780 Bochum, Germany
\and 
        European Southern Observatory, Karl-Schwarzschildstr. 2, 
        D-85748 Garching b. M\"unchen, Germany
\and 
        IPAC, California Institute of Technology (Caltech), Pasadena, 
        CA 91125, USA
\and
        Max-Planck-Institut f\"ur Radioastronomie, Auf dem H\"ugel 69,
        Postfach 2024, D-53010 Bonn, Germany
}
\offprints{Martin Haas, haas@astro.rub.de} \date{xx.07.2005/xxx}
\authorrunning{M. Haas et al.}
\titlerunning{Spitzer IRS spectroscopy of 3CR radio galaxies and
        quasars}

\abstract{With the Spitzer Space Telescope Infrared Spectrograph (IRS)  
we have observed seven {\it powerful} FR\,2 radio galaxies and seven
quasars. Both samples, the galaxies and the quasars, 
are comparable in isotropic 178 MHz luminosity
(10$^{\rm 26.5}$ W/Hz $\simless$ P$_{\rm 178\,MHz}$ $\simless$ 10$^{\rm 29.5}$ W/Hz)
and in redshift range ($0.05 \simless z \simless 1.5$). 
We find for both samples similar distributions in the luminosity ratios of high- to
low-excitation lines 
([NeV]$_{\rm 14.3 \mu m}$\,/\,[NeII]$_{\rm 12.8 \mu m}$) and of
high-excitation line to radio power 
([NeV]$_{\rm 24.3 \mu m}$\,/\,P$_{\rm 178\,MHz}$).
This solves the long debate about the apparent
difference of quasars and radio galaxies in favor of the
orientation-dependent unified schemes.
Furthermore, the luminosity ratio  
[OIII]$_{\rm 500.7 nm}$\,/\,[OIV]$_{\rm 25.9 \mu m}$ 
of most galaxies is by a factor of ten 
lower than that of the quasars. This suggests 
that the optical emission from the
central NLR is essentially absorbed (A$_{\rm V}$\,$\simgreat$\,3)
in the {\it powerful}  FR\,2 galaxies and that the [OIII]$_{\rm 500.7 nm}$ luminosity 
does not serve as isotropic tracer for testing the unified schemes.  

\keywords{      Galaxies: active --
                Galaxies: nuclei --
		Galaxies: quasars: general --  Infrared: galaxies 
         }	}

\maketitle

\section{Introduction}

According to the orientation-dependent AGN unified schemes both the 
powerful FR\,2 radio galaxies and the quasars are believed to belong to the
same parent population (Barthel 1989) with their observed properties depending
on our viewing angle towards the obscuring dust torus (Antonucci 1993).  
First evidence for this hypothesis was provided by
spectro-polarimetric observations in those  galaxies with "edge-on" torus,
  where scattering particles located above and below the dust torus
allow for viewing the AGN-typical broad spectral lines of the central region
(e.g. Antonucci 1984).
In general 
a proper test of the unification is best
performed by observations of isotropic
quantities, which are not affected by orientation-dependent extinction
or beaming. Since the fraction of the
nuclear luminosity that is absorbed by dust must be re-radiated 
in the infrared, the optically thin
far-infrared
  (FIR, 40-1000\,$\mu$m)
  emission is such an isotropic measure. 
From sensitive photometry with the ISO satellite one found similar isotropic 
FIR dust luminosities for radio galaxies and quasars 
from the 3CR catalogue, after normalisation by their likewise isotropic 
178 MHz radio power (Meisenheimer et al. 2001, Haas et al. 2004). 
The distributions showed a large dispersion, which may be 
caused by the
various evolutionary stages of these complex sources, and it is still under
debate whether the FIR luminosity arises from starbursts rather than from 
the AGN itself. 

Both the radio galaxies and the 
quasars are luminous mid-infrared
  (MIR, 10-40\,$\mu$m)
emitters, and
the mid- to far-infrared luminosity ratio of the 3CR sources  
exceeds that of pure starburst galaxies; but 
the radio galaxies show a lower L$_{\rm MIR}$ / L$_{\rm FIR}$ than
quasars (Haas et al. 2004, Siebenmorgen et al. 2004). 
Either the emission from the putative dust torus is optically thick
even at MIR wavelengths (Pier \& Krolik 1993) - hence not isotropic -, 
or the intrinsic AGN luminosities are different supporting  modifications of
the orientation-dependent unification such as the receding torus model
(Lawrence 1991).
Further objections against the pure orientation-dependent unification come 
from the lower [OIII]$_{\rm 500.7 nm}$ / P$_{\rm 178\,MHz}$ luminosity ratios
of galaxies, assuming that {\it most} of the [OIII]$_{\rm 500.7 nm}$ emission arises from
the {\it extended} narrow line region not affected by central extinction.

\begin{figure*}
\hbox{\hspace{0cm}
\psfig{file=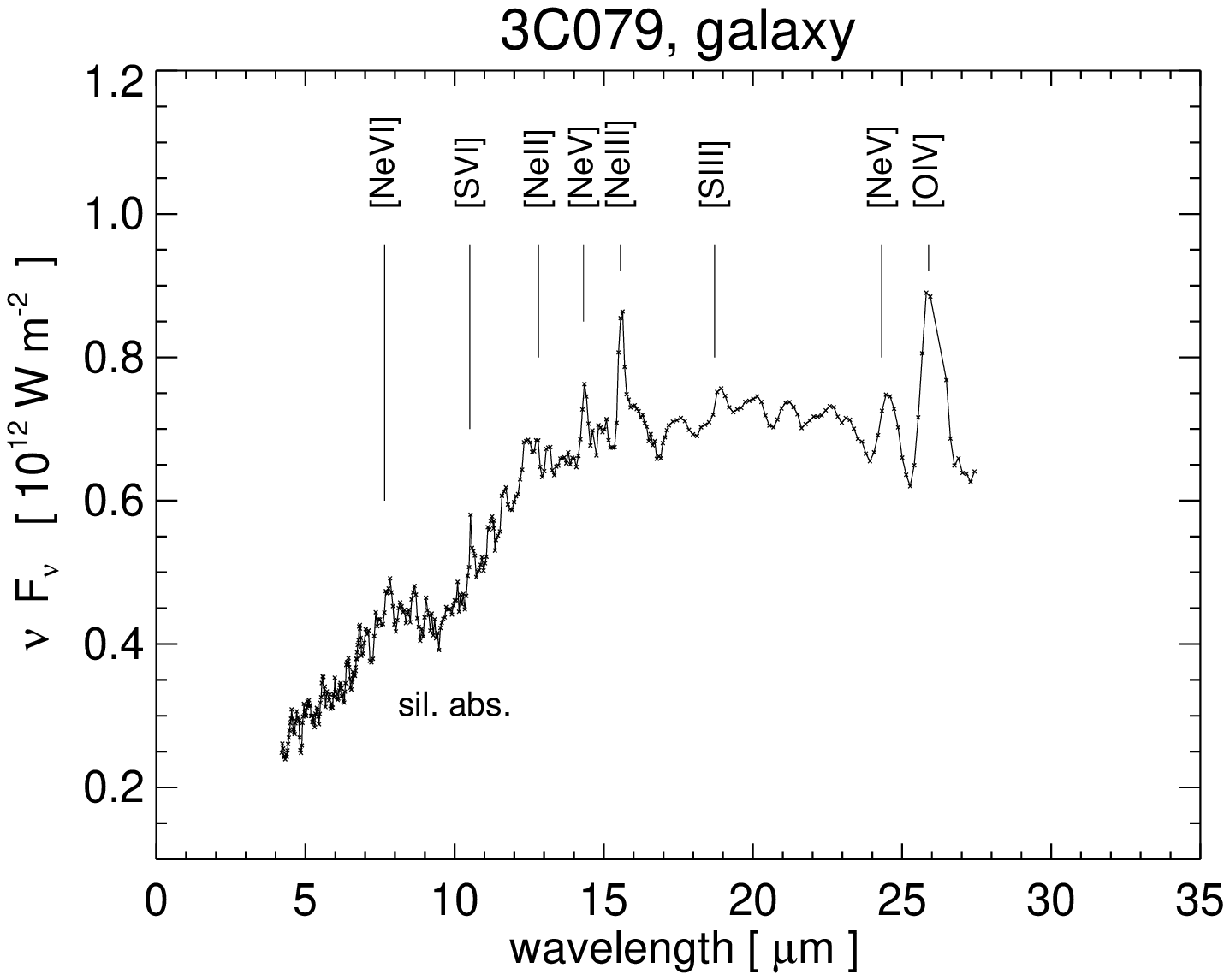,width=6cm,clip=true}
\hspace{0.cm}
\psfig{file=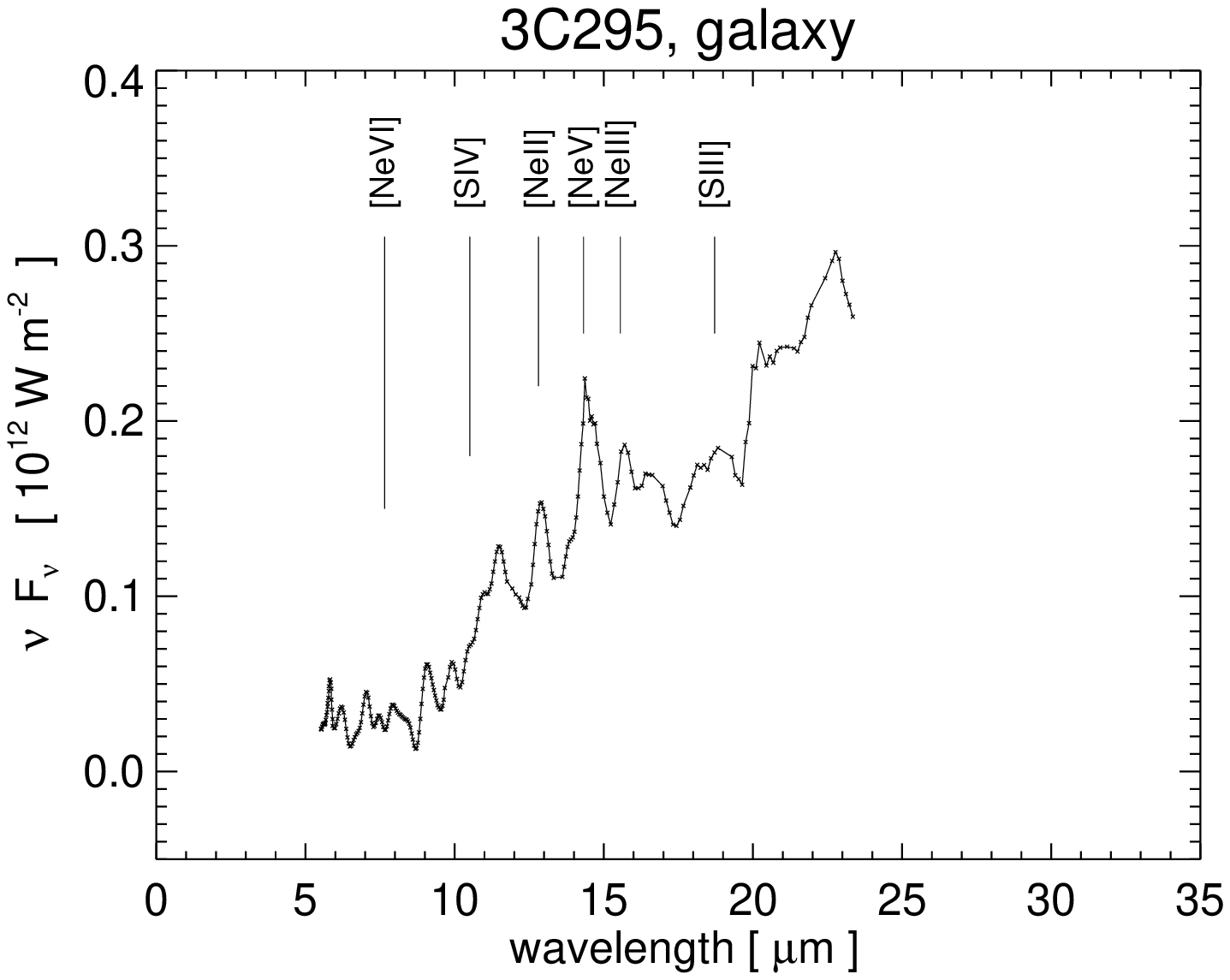,width=6cm,clip=true}
\hspace{0.cm}
\psfig{file=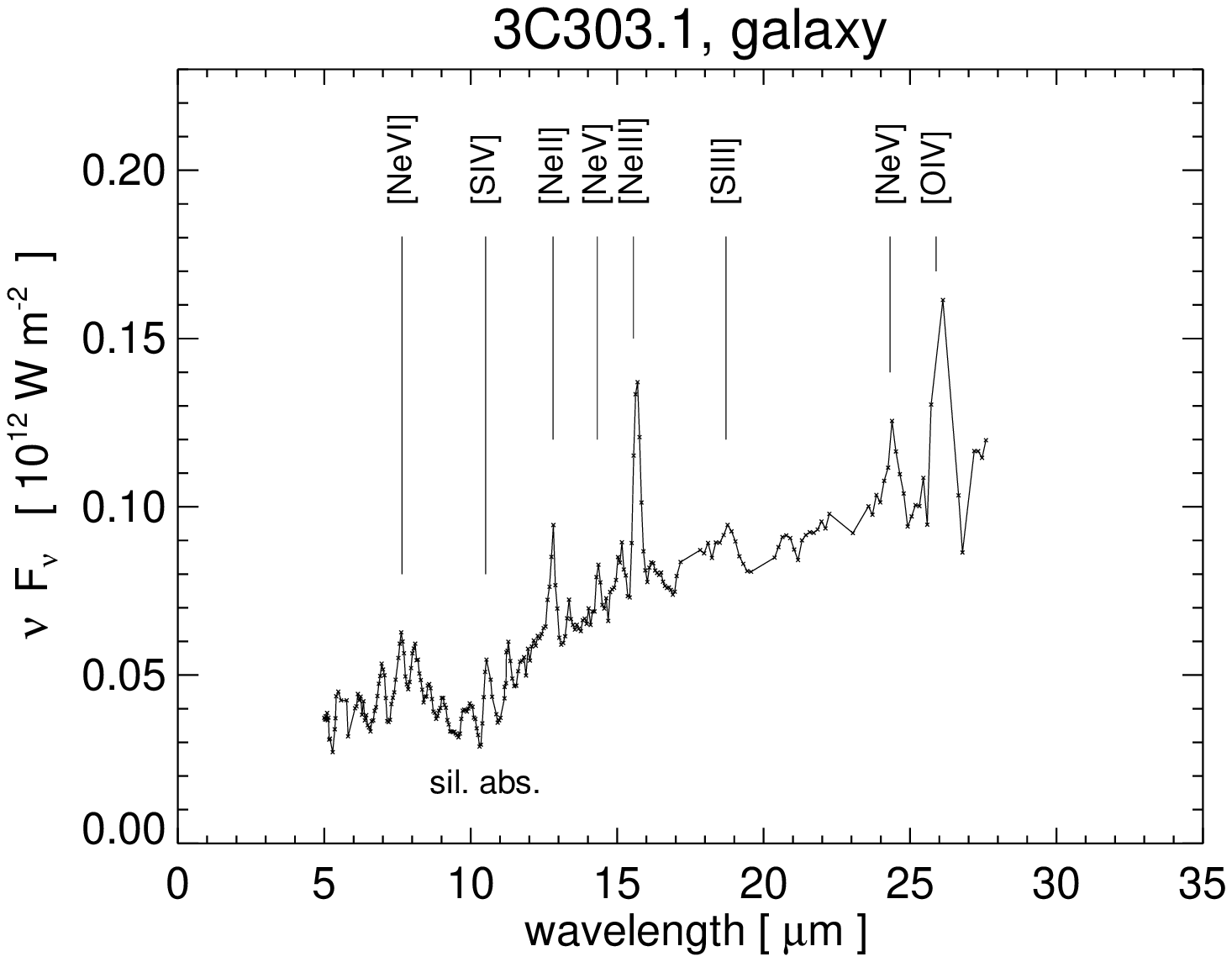,width=6cm,clip=true}
}
\hbox{\hspace{0cm}
\psfig{file=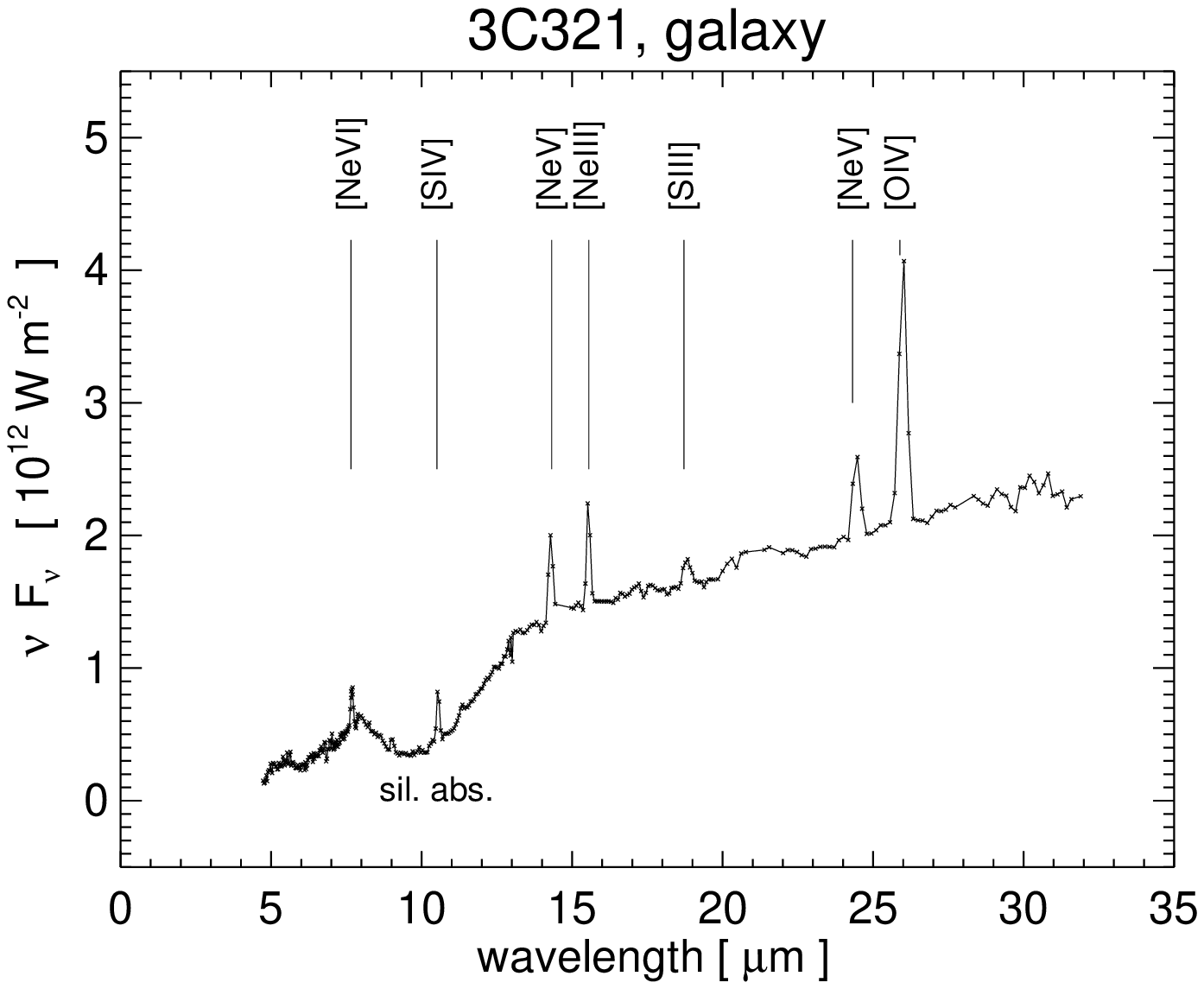,width=6cm,clip=true}
\hspace{0.cm}
\psfig{file=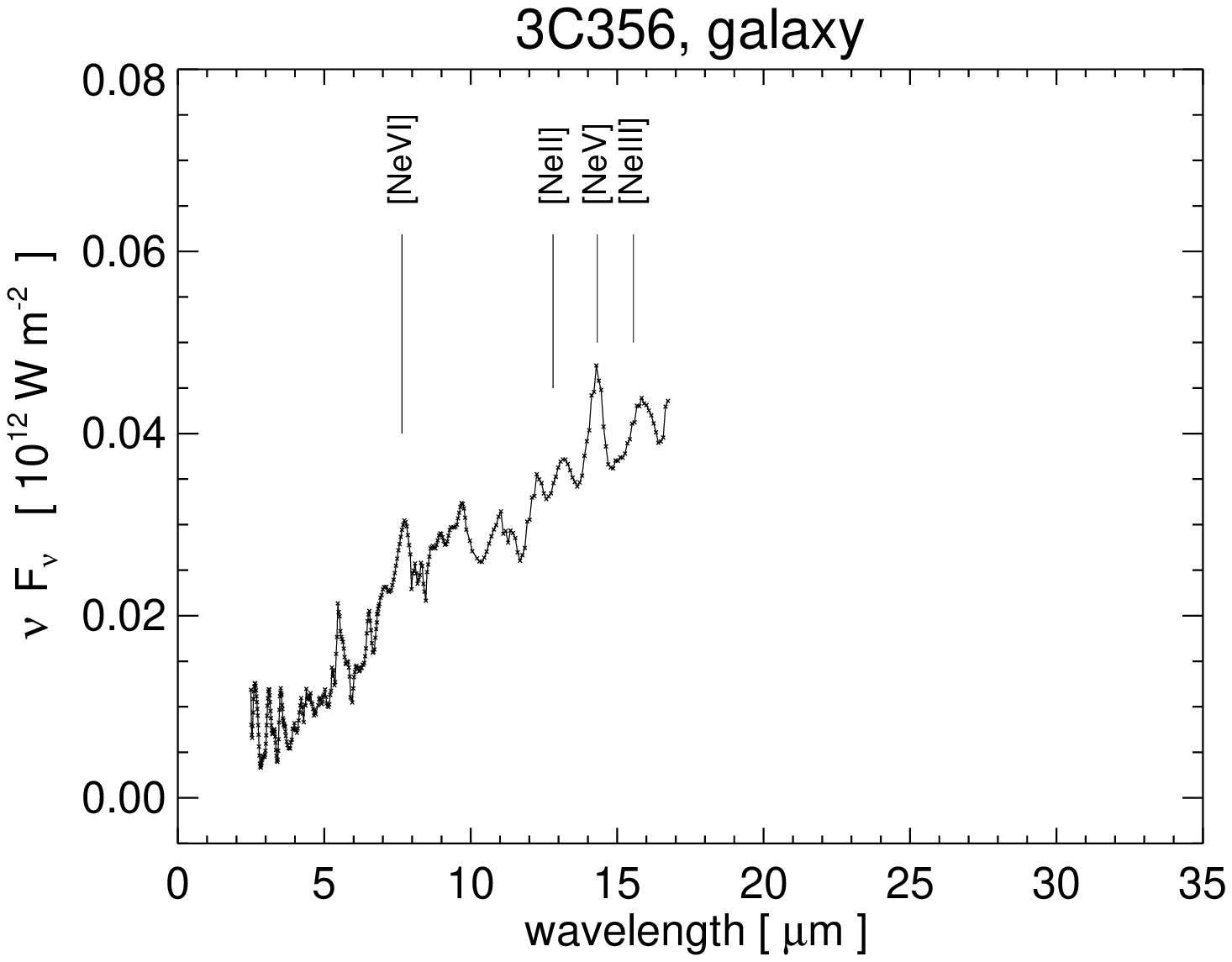,width=6cm,clip=true}
\hspace{0.cm}
\psfig{file=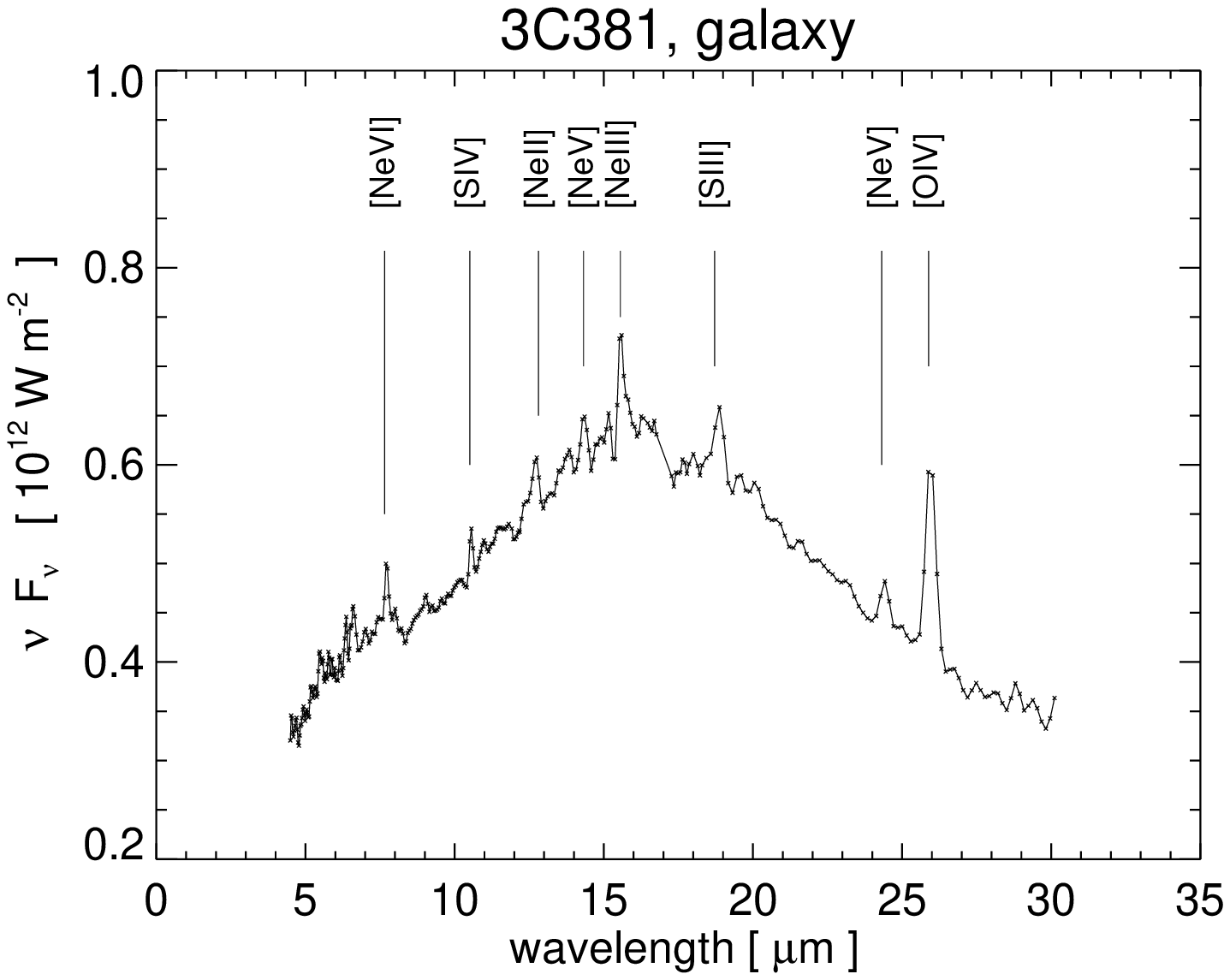,width=6cm,clip=true}
}
\hbox{\hspace{0cm}
\psfig{file=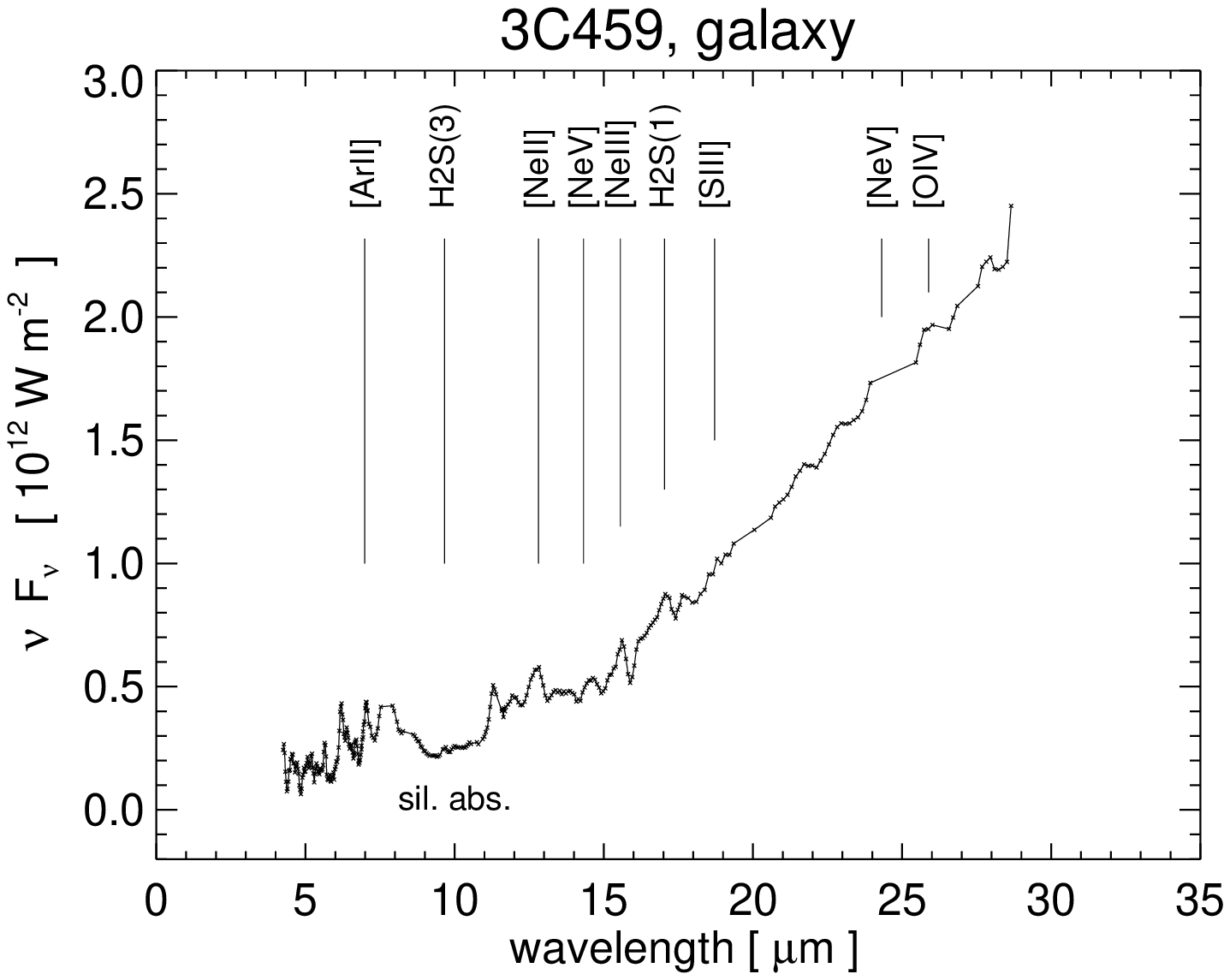,width=6cm,clip=true}
\hspace{0.cm}
\psfig{file=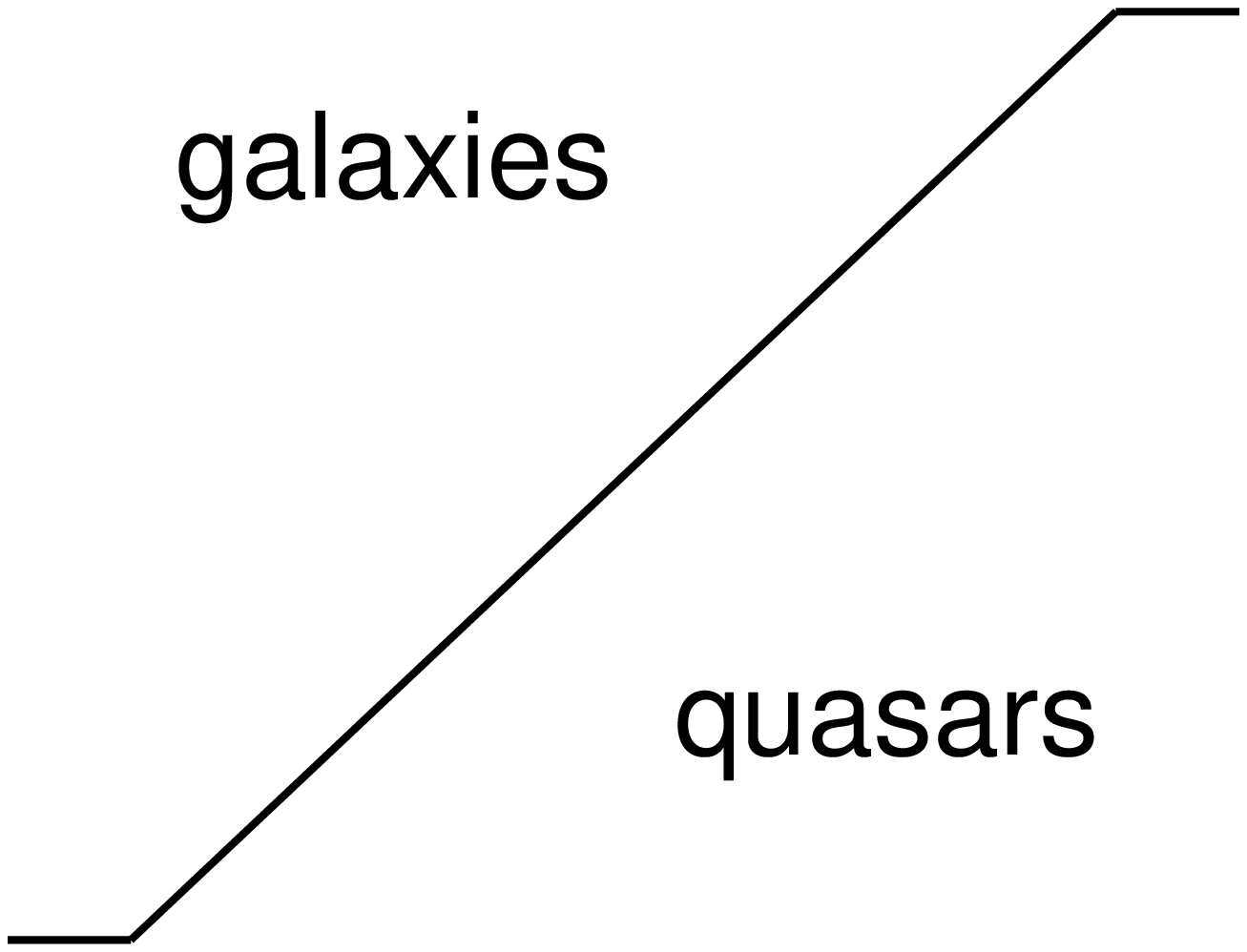,width=6cm,clip=true}
\hspace{0.cm}
\psfig{file=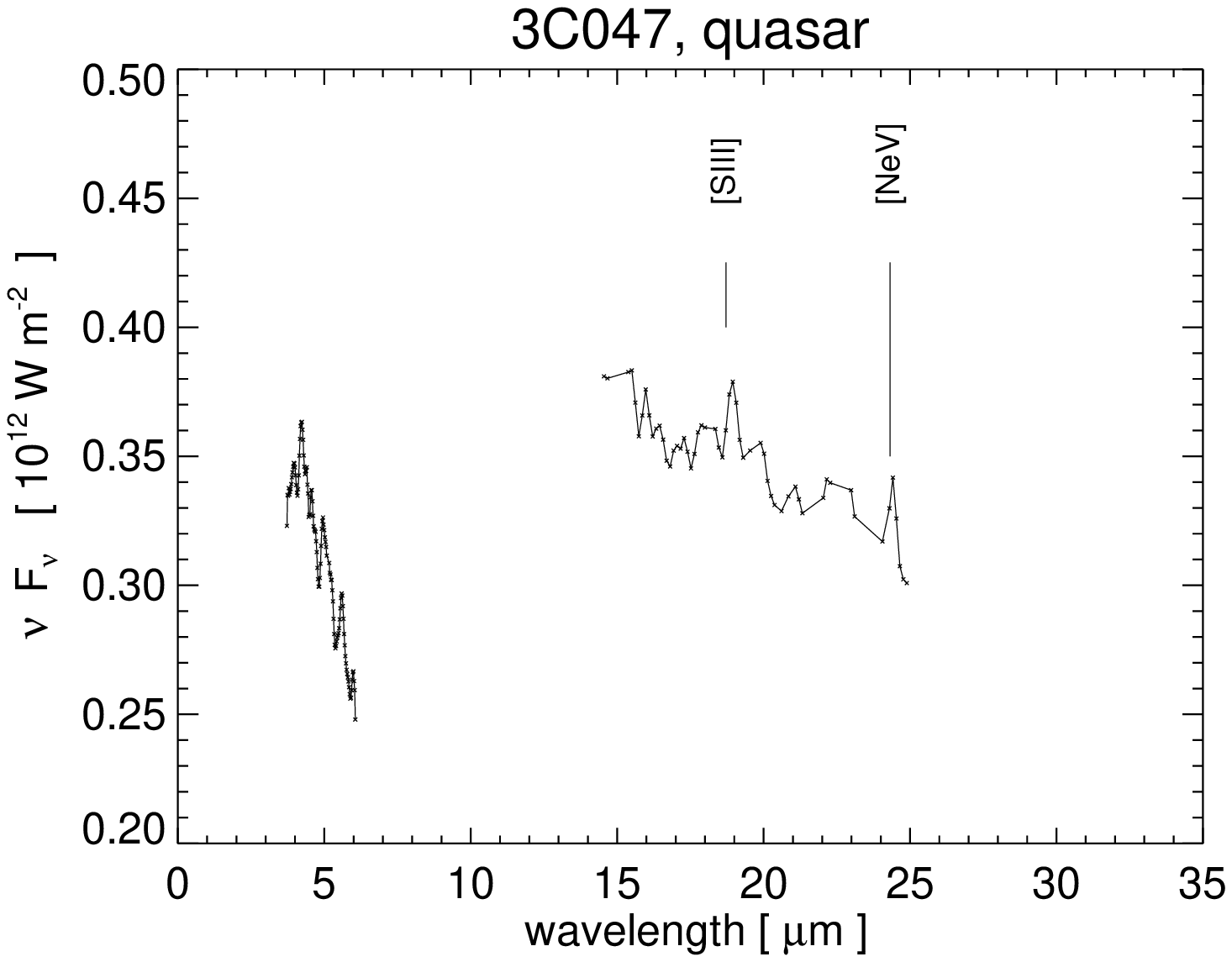,width=6cm,clip=true}
}
\hbox{\hspace{0cm}
\psfig{file=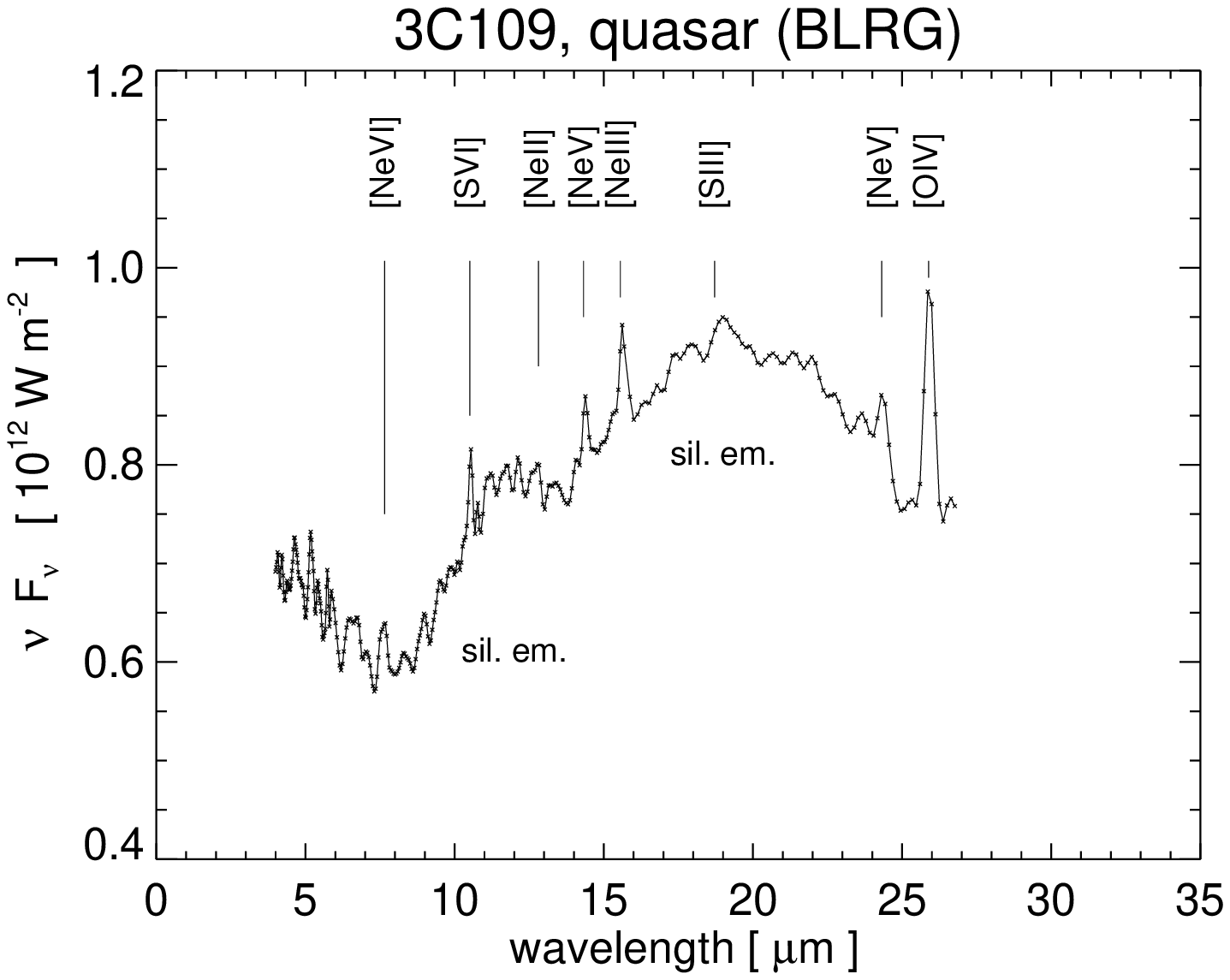,width=6cm,clip=true}
\hspace{0.cm}
\psfig{file=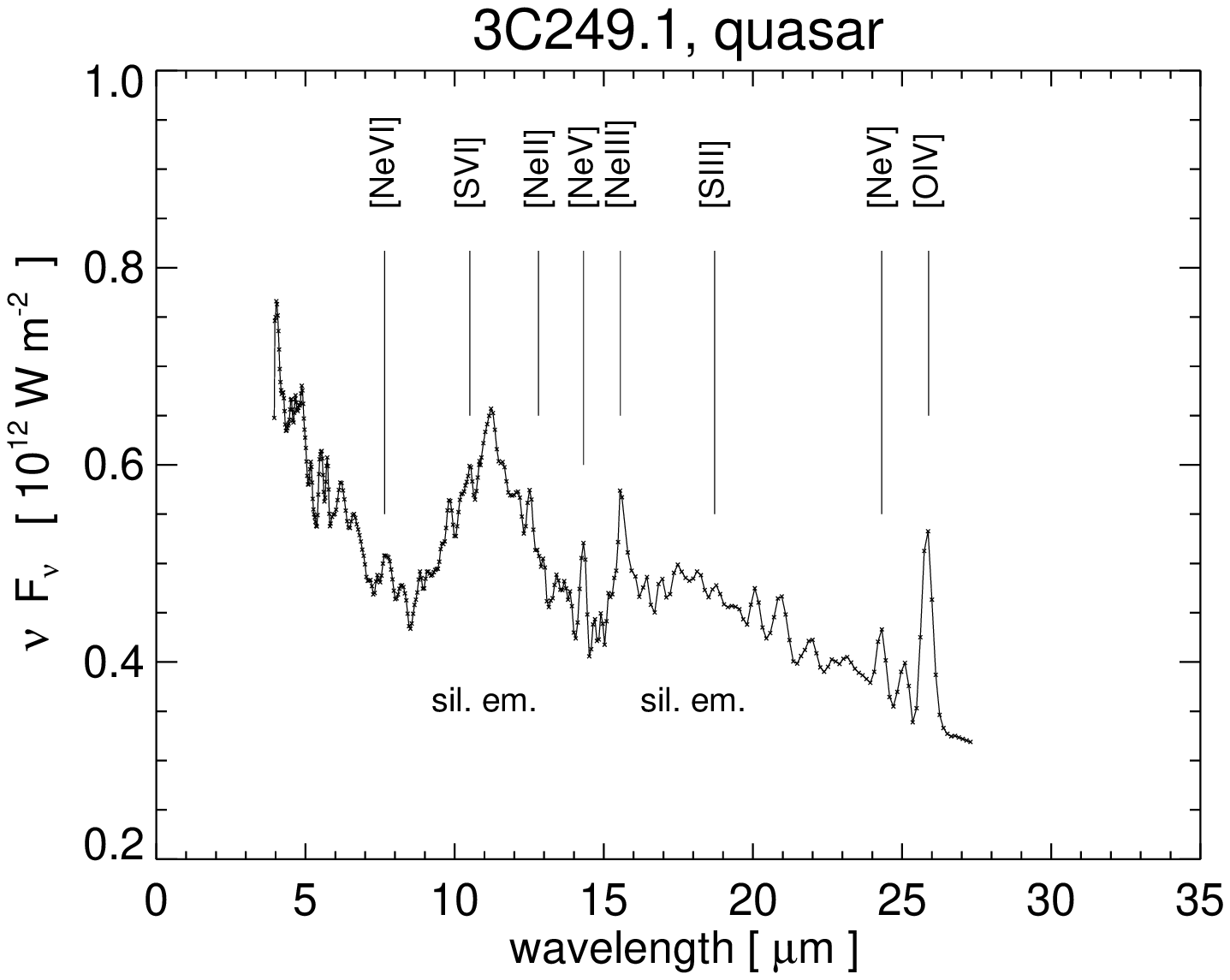,width=6cm,clip=true}
\hspace{0.cm}
\psfig{file=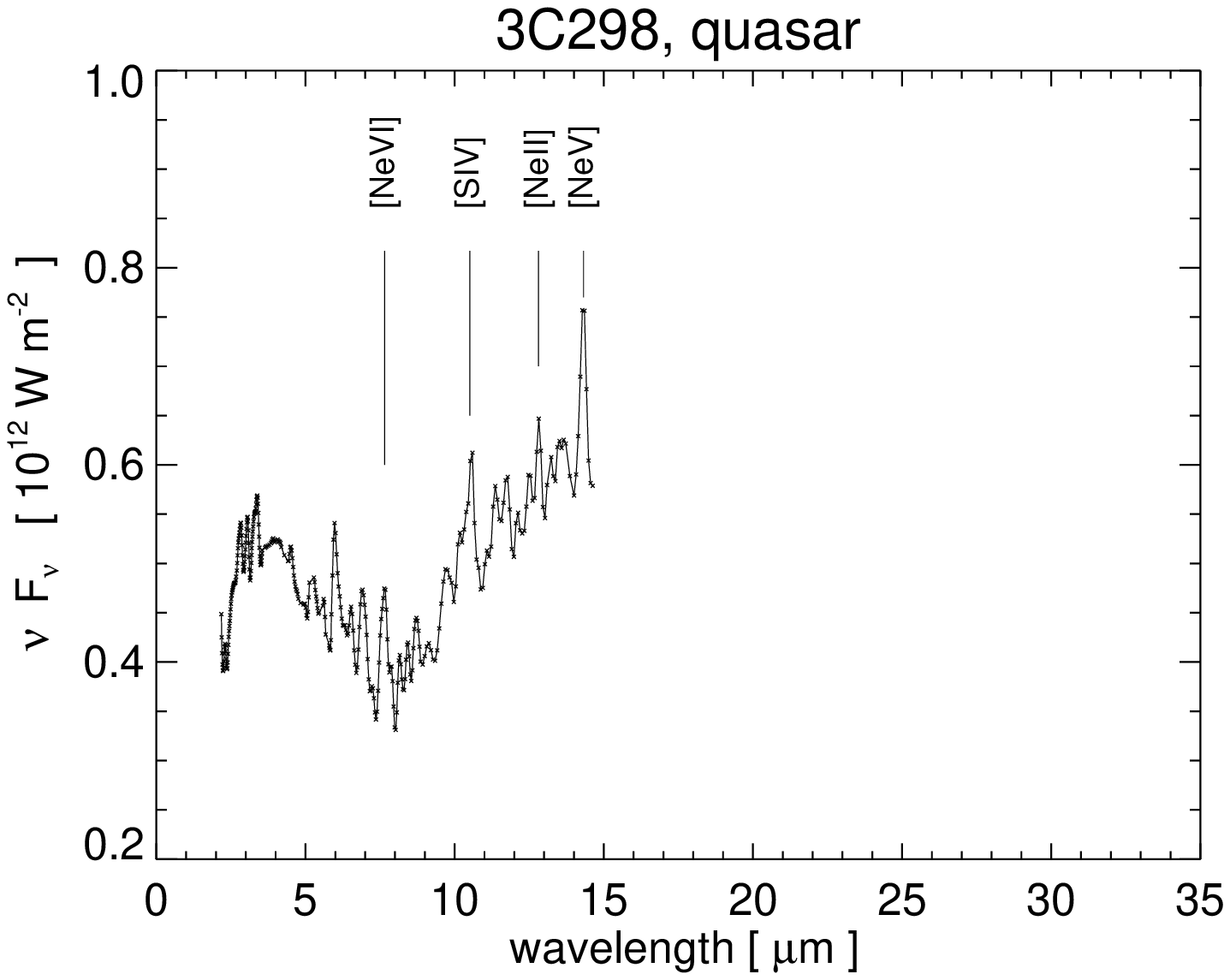,width=6cm,clip=true}
}
\hbox{\hspace{0cm}
\psfig{file=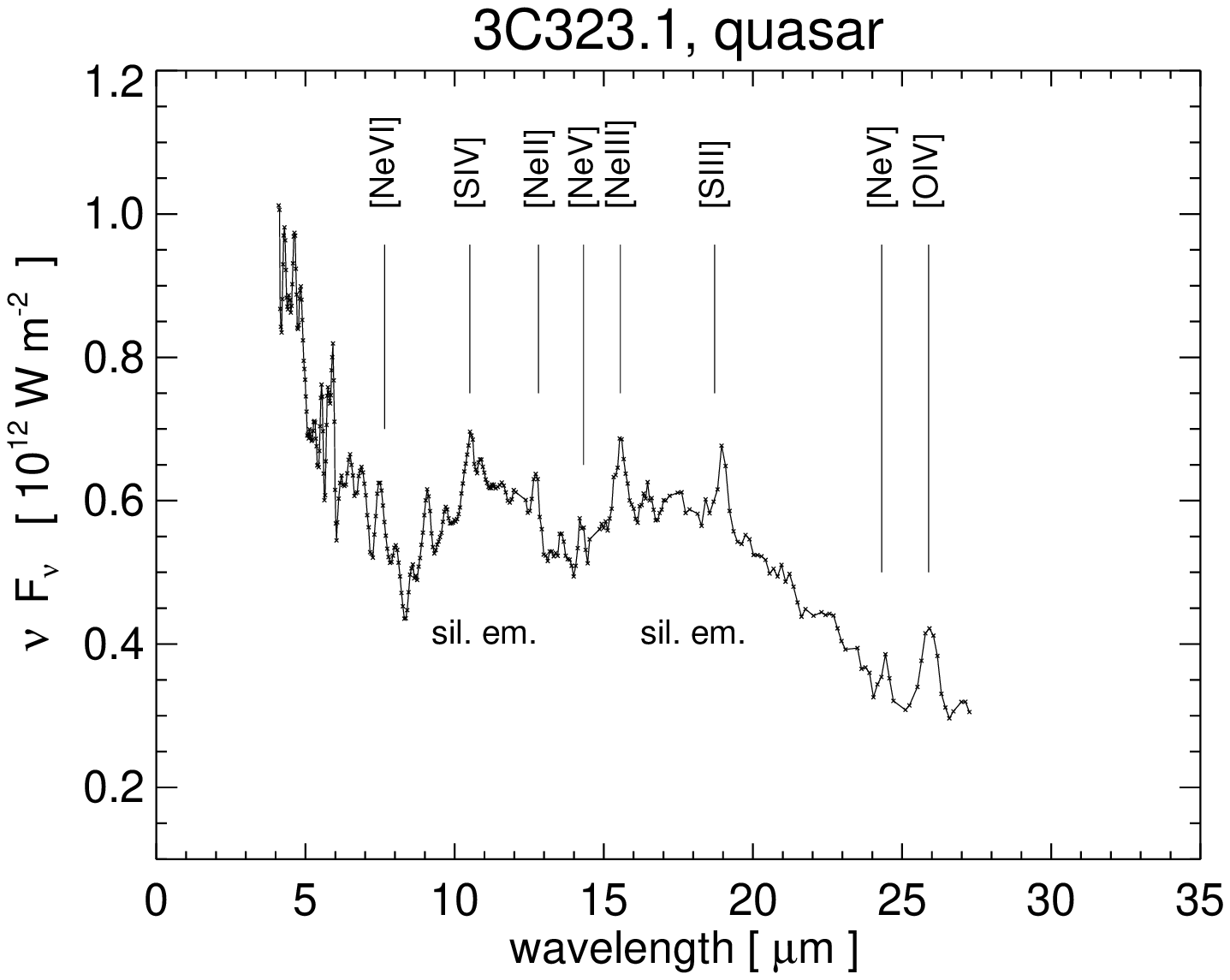,width=6cm,clip=true}
\hspace{0.cm}
\psfig{file=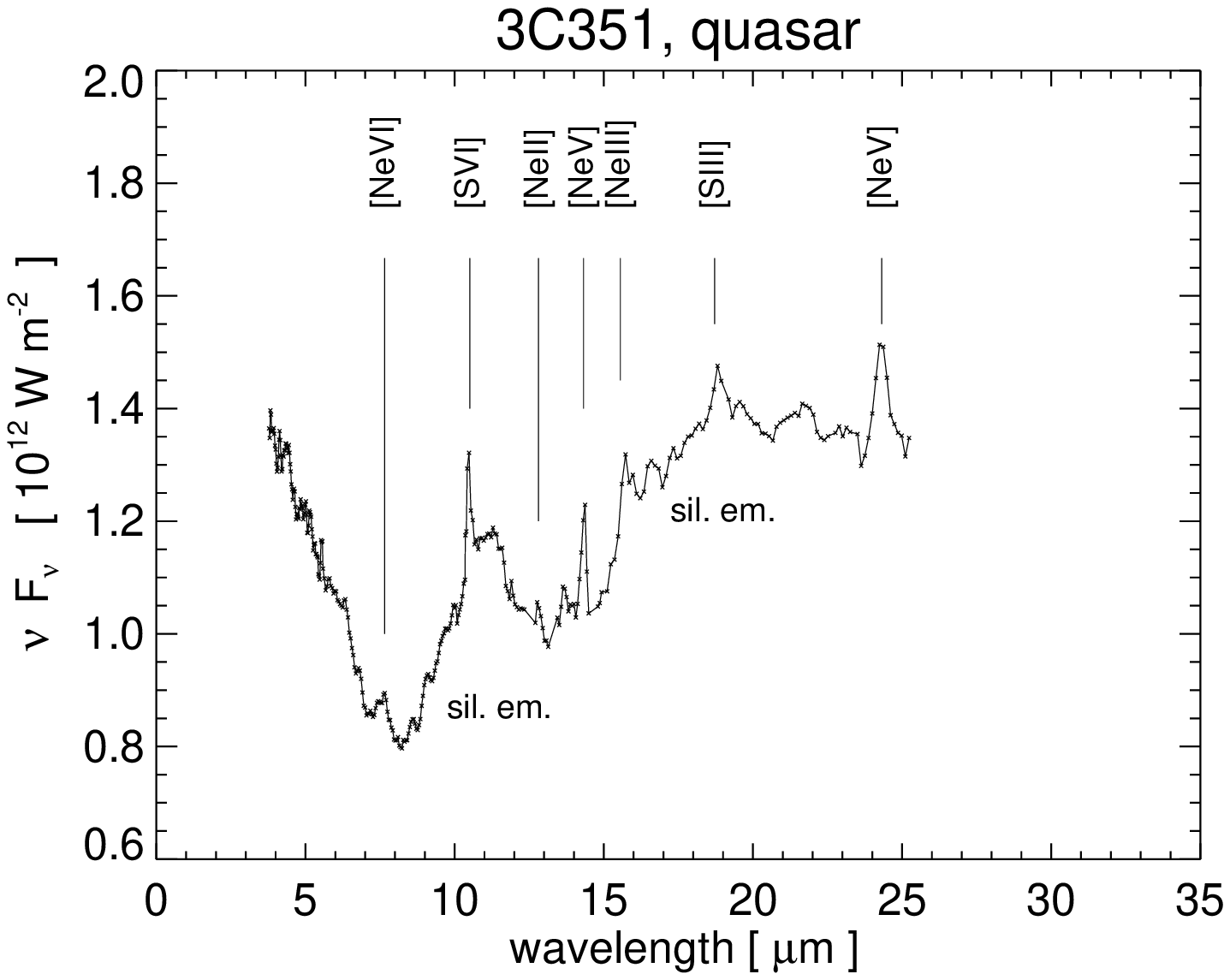,width=6cm,clip=true}
\hspace{0.cm}
\psfig{file=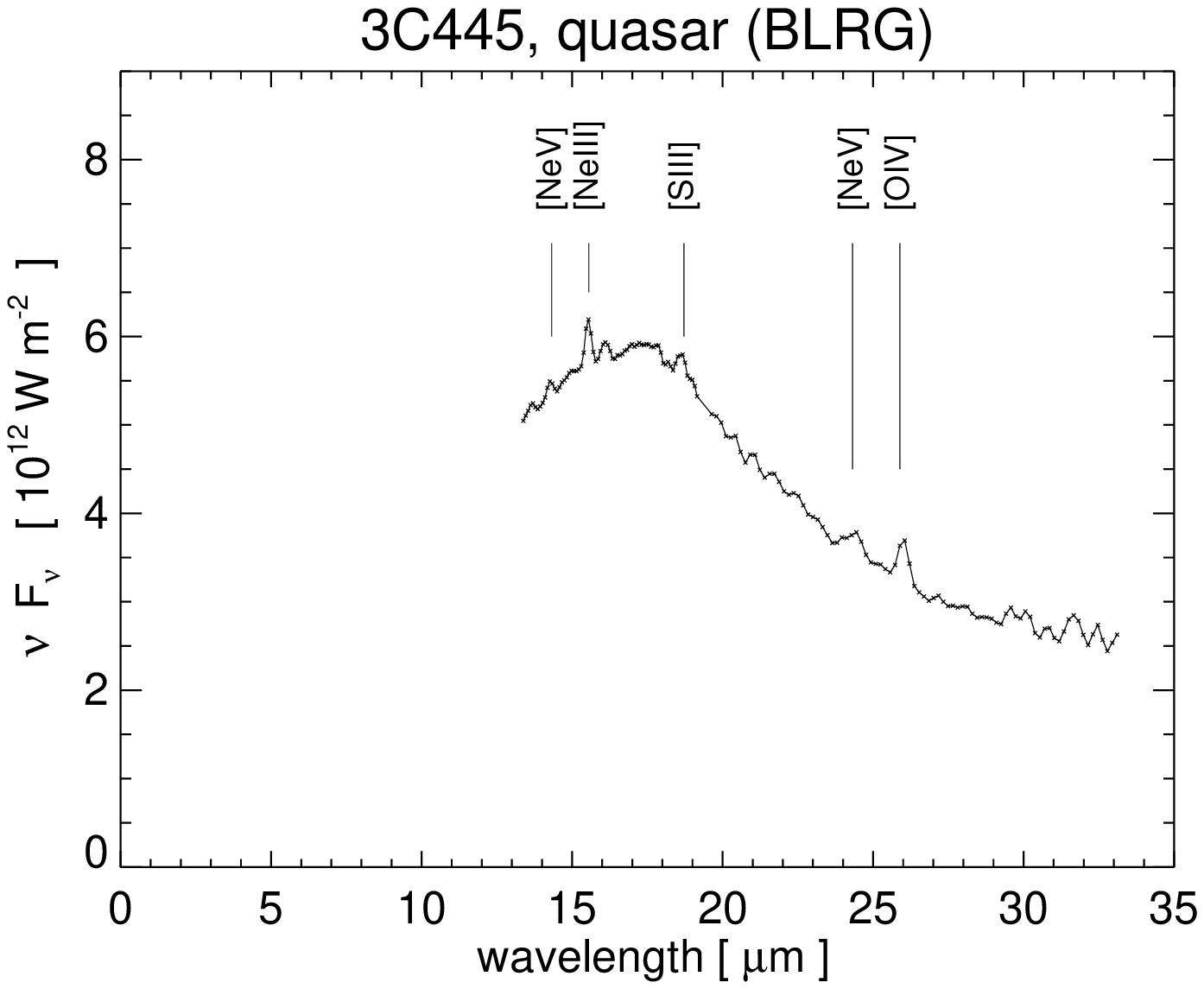,width=6cm,clip=true}
}
\caption {
{ IRS Spectra of the radio galaxies {\it (top panels)} and quasars {\it (bottom panels)},
  shifted to the object rest frames.
  The important
  emission lines as well as the silicate absorption and emission
  features are marked.
  For 3C\,047 and 3C\,445 only two of the four
  IRS channels were  assigned to our program. 
}
\label {fig1}
}
\end{figure*}

\begin{figure}
\hbox{
\psfig{file=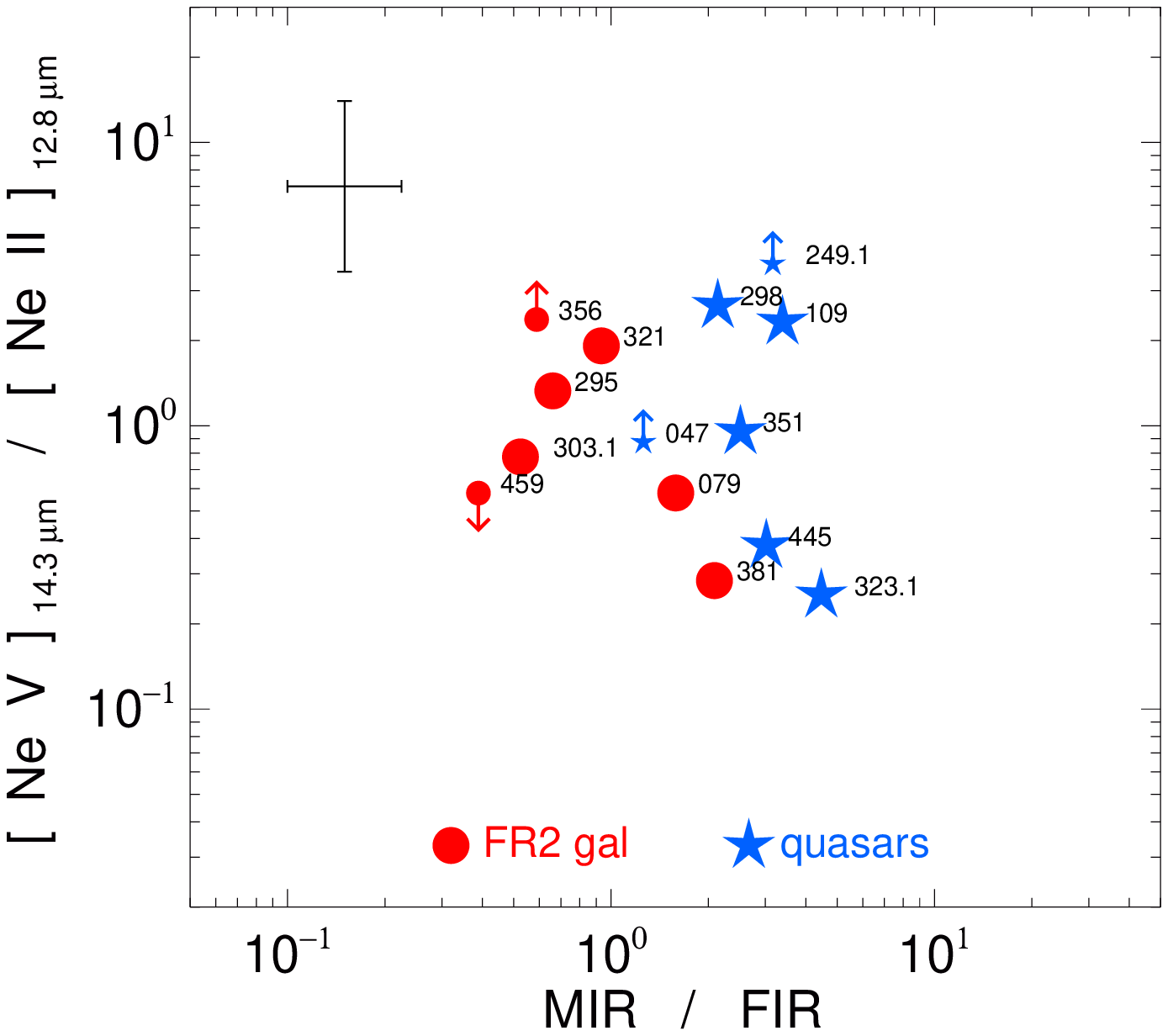,width=9cm,clip=true}
}
\caption{
  {Luminosity ratios of high over low excitation lines versus mid-to
  far-infrared luminosity.
  For 3C\,047, 3C\,079, 3C\,321 and 3C\,445) [Ne\,II] was substituted by [S\,III],
  and for 3C\,047 [Ne\,V]$_{\rm 14.3 \mu m}$
  by [Ne\,V]$_{\rm 24.3 \mu m}$.
  The cross indicates errors of a factor 1.5 (MIR/FIR) and 2 ([Ne\,V]/[Ne\,II]).
  }
  \label {fig2}
}
\end{figure}

The high- and low-excitation 
MIR spectral lines suffer from much less extinction than optical ones. 
Therefore, we observed 7 
quasars (and BLRGs) and 7 
radio galaxies from the 3CR catalogue. 
First results from this sample 
on the discovery of 10 $\mu$m silicate emission in
quasars have already been presented by Siebenmorgen et al. (2005). Here, we
focus on the investigation of the emission lines.

\section{Sample selection and Observations}

  The steep radio spectrum quasars and powerful galaxies were randomly
  selected from the 3CR catalogue, 
  so that the samples match in 178 MHz luminosity as well as in
  redshift.
  Note that we do not consider FR\,1 or low-power FR\,2 sources, since for them
  the orientation-dependent unification picture may not apply.
  Thus, any conclusions drawn here refer to FR\,2 sources with
  10$^{\rm 26.5}$ W/Hz $\simless$ P$_{\rm 178\,MHz}$ $\simless$ 10$^{\rm 29.5}$ W/Hz
  and $0.05 \simless z \simless 1.5$. 

  More specifically, the sources were selected from 
  that sub-sample of the 3CR catalogue
  which (1) was not reserved by guarantee time observations
  and (2) was observed by ISO with sufficient S/N (Haas et al. 2004, Siebenmorgen et al. 2004)
  in order to provide the flux estimates
  required for carefully planning the Spitzer IRS observations.
  This "ISO subset" provides a fairly representative sub-sample of the 3CR catalogue,
  since the 
  objects varied widely in apparent brightness
  and the selection criteria for the sources
  were low cirrus foreground and good visibility to the satellite.
  Thus, we do not expect that the targets selected for
  our Spitzer IRS observations are biased in favour of
  MIR-loud 3CR sources.

The objects were observed between 5 and 35 $\mu$m
with the infrared spectrograph (IRS, Houck et al. 2004) of the Spitzer Space
Telescope (Werner et al. 2004)  
in the two IRS low-resolution
(64 $<$ $\lambda$ / $\Delta$$\lambda$ $<$ 128) modules in staring mode. Our
analysis starts with the two dimensional data frames from the Spitzer pipeline
(Higdon et al. 2004) using the latest calibration files.  At this point, the
major instrumental effects have already been removed.  We subtracted the sky
background, using pairs of frames, where the source appears at two different
positions along the spectrometer slit.  We interactively extracted the one
dimensional spectra.  Before averaging, parts with low reproducibility between
integrations were discarded.
 
\section{Results and Discussion}

\begin{figure}
\hbox{
\psfig{file=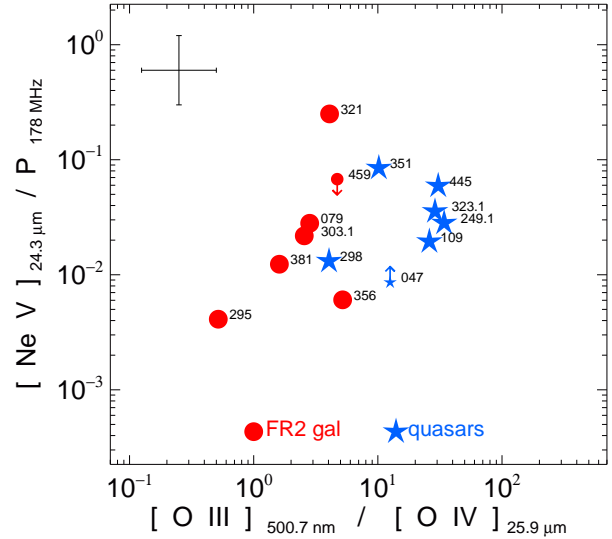,width=9cm,clip=true}
}
\caption{
  {High-excitation line to radio luminosities 
  versus [OIII]$_{\rm 500.7 nm}$\,/\,[OIV]$_{\rm 25.9 \mu m}$
  luminosities. 
  For 3C\,295, 3C\,298 and 3C\,356 [Ne\,V]$_{\rm 24.3 \mu m}$
  was substituted by [Ne\,V]$_{\rm 14.3 \mu m}$,
  and for
  3C\,047, 3C\,295, 3C\,298, 3C\,351 and 3C\,356 [O\,IV]$_{\rm 25.9 \mu m}$
  by 3\,$\times$\,[Ne\,V]. The cross indicates errors of a factor 2.
  }
  \label {fig3}
}
\end{figure}

Figure~\ref{fig1} depicts the IRS spectra shifted to the object rest frames.  
The spectra show numerous emission lines as superposed on diverse continuum
shapes, caused by various hot and warm dust components, and the broad silicate 
features around 10 and 18 $\mu$m.
The silicate features are typically seen in absorption for the galaxies
(e.g. in 3C\,079, 3C\,303.1, 3C\,321, 3C\,459) and
in emission for the quasars (e.g. in 3C\,109, 3C\,249.1, 3C\,323.1,
3C\,351).
Also
in the NIR (3-10\,$\mu$m) 
shortward of 7\,$\mu$m the continuum shows a typical difference
between galaxies and quasars: in quasars 
it is steeply rising with decreasing wavelength while in galaxies it is
flat or declining.
  From both the different silicate features and the bluer NIR continuum of quasars 
we 
suggest that the reported difference of the L$_{\rm MIR}$/L$_{\rm FIR}$ ratio
(L$_{\rm 10-40 \mu m}$/L$_{\rm 40-1000 \mu m}$)
is in fact caused by 
suppression of the MIR continuum in
galaxies and by its enhancement in quasars. 

\begin{table*}
  {
    \caption{ Sample parameters. The types are abbreviated as Q=quasar, B = broad line radio galaxy, N = narrow line radio galaxy.
      The luminosity distance D$_{\rm L}$ is computed adopting a $\Lambda$-cosmology with
      H$_{\circ}$ = 70 km s$^{\rm -1}$Mpc$^{\rm -1}$, $\Omega$$_{\rm m}$ = 0.27 and $\Omega$$_{\rm \lambda}$ = 0.73.
      Below the line wavelength we list the ionisation potential
      needed to create the species from the next lower ionisation stage.
      The line luminosities are given logarithmically in units of [W].
      The [OIII]$_{\rm 500.7 nm}$ luminosities are taken from the compilation by Grimes et al. (2004)
      and from Boroson \& Oke (1984), Tadhunter et al. (1993),
      Mc Carthy et al. (1995) and
      Hirst et al. (2003).
    }
    \label {tab1}   }
  {\scriptsize
    \begin{tabular}{@{\hspace{1.1mm}}l|c@{\hspace{1.1mm}}rr|rrrr rrrr|rr@{\hspace{1.1mm}}}
      \hline
object& type& z~~ & D$_{\rm L}$ &[Ne\,VI]&[S\,IV]&[Ne\,II]&[Ne\,V]&[Ne\,III]&[S\,III]&[Ne\,V]&[O\,IV]&[O\,III]& P$_{\rm 178 MHz}$      \\
      &     &     & Mpc &7.6\,$\mu$m&10.5\,$\mu$m&12.8\,$\mu$m&14.3\,$\mu$m&15.5\,$\mu$m&18.7\,$\mu$m&24.3\,$\mu$m&25.9\,$\mu$m&500.7\,nm& \\
      &     &     &             & 126.2\,eV& 34.8\,eV& 21.6\,eV & 97.1\,eV& 41.0\,eV  & 23.3\,eV & 97.1\,eV& 54.9\,eV& 35.1\,eV &            \\
      \hline
  3C\,079   &N&   0.256&   1283 &$<$34.02&   34.19 &$<$34.12 &   34.36 &   34.60 &   34.37 &   34.53 &   35.14 & 35.59 &  36.08 \\
  3C\,295   &N&   0.461&   2559 &$<$34.28&$<$34.33 &   34.64 &   34.76 &   34.62 &$<$34.78 &         &         & 34.99 &  37.15 \\
  3C\,303.1 &N&   0.267&   1347 &$<$33.76&   34.04 &   33.92 &   33.65 &   34.22 &   33.53 &   33.89 &   34.31 & 34.71 &  35.55 \\
  3C\,321   &N&   0.096&    435 &   33.77&   33.90 &$<$33.43\parbox{0cm}{$^{\rm a}$} &   34.19 &   34.11 &   33.73 &   34.16 &   34.64 & 35.25 &  34.76 \\
  3C\,356   &N&   1.079&   7296 &   34.74&         &$<$34.68 &   35.05 &$<$34.49 &         &         &         & 36.28 &  37.27 \\
  3C\,381   &N&   0.160&    758 &   33.74&   33.41 &   33.62 &   33.60 &   33.96 &   33.92 &   33.43 &   34.19 & 34.40 &  35.34 \\
  3C\,459   &N&   0.220&   1081 &        &         &   34.51 &$<$34.36 &   34.30 &$<$34.02 &$<$34.51 &   34.47 & 35.14 &  35.85 \\
\hline
  3C\,047   &Q&   0.425&   2322 &        &         &         &         &         &   34.53 &$>$34.67\parbox{0cm}{$^{\rm b}$} &         & 36.28 &  36.56 \\
  3C\,109   &B&   0.306&   1577 &   34.61&   34.50 &   33.94 &   34.30 &   34.52 &$<$33.89 &   34.41 &   34.91 & 36.32 &  36.12 \\
  3C\,249.1 &Q&   0.311&   1607 &$<$34.57&         &$<$33.98 &   34.46 &   34.63 &$<$34.00 &   34.28 &   34.86 & 36.39 &  35.83 \\
  3C\,298   &Q&   1.436&  10427 &   36.19&   36.29 &   35.93 &   36.36 &         &         &         &         & 37.30\parbox{0cm}{$^{\rm c}$} &  38.24 \\
  3C\,323.1 &Q&   0.264&   1329 &$<$34.35&   34.51 &   34.39 &   34.13 &   34.38 &   34.49 &   34.10 &   34.56 & 36.02 &  35.61 \\
  3C\,351   &Q&   0.371&   1975 &   34.67&         &   34.66 &   35.01 &   34.99 &   34.80 &   35.06 &         & 36.46 &  36.13 \\
  3C\,445   &B&   0.056&    247 &        &         &         &   33.05 &   33.56 &   33.32 &   33.29 &   33.63 & 35.11 &  34.52 \\
    \hline
    \end{tabular}
    $^{\rm a}$ line between two channels, $^{\rm b}$ line on edge of channel,\\
    $^{\rm c}$ adopting L([OIII]) = 0.1 $\times$ L(total H$_{\alpha}$) from the mean SDSS quasar template (Vanden Berk et al. 2005)
  }
\end{table*}

The spectra show AGN-typical high-excitation lines like 
[Ne\,V]$_{\rm 14.3 \mu m}$, [Ne\,V]$_{\rm 24.3 \mu m}$ and
[O\,IV]$_{\rm 25.9 \mu m}$, as well as low-excitation lines like 
[Ne\,II]$_{\rm 12.8 \mu m}$ and [S\,III]$_{\rm 18.7 \mu m}$. 
Table\,\ref{tab1} lists the 
line luminosities. 
We estimate the uncertainty of  
the line fluxes from the subtraction of the underlying noisy continuum
to $\la$\,40\%; combined with the
absolute flux calibration uncertainty of $\la$\,20\% this results in a
total line flux uncertainty of $\sim$\,50\% which 
is sufficient for the conclusions drawn here.
In order to compare the two samples by means of representative lines, 
when such a line is unobserved or poorly measured, we substitute it
by another one of similar (or higher) ionisation potential with
appropriately scaled strength determined from average line ratios: 
[Ne\,II]$_{\rm 12.8 \mu m}$ by  [S\,III]$_{\rm 18.7 \mu m}$, 
[Ne\,V]$_{\rm 24.3 \mu m}$ by [Ne\,V]$_{\rm 14.3 \mu m}$ (and vice versa),
and [O\,IV]$_{\rm 25.9 \mu m}$ by 3\,$\times$\,[Ne\,V]$_{\rm 24.3 \mu m}$.
This leaves the scatter in the line ratios below a factor of two.

Figure~\ref{fig2} displays 
the 
[NeV]$_{\rm 14.3 \mu m}$\,/\,[NeII]$_{\rm 12.8 \mu m}$ luminosity ratio
versus L$_{\rm MIR}$\,/\,L$_{\rm FIR}$. 
While the radio galaxies show on average a lower 
L$_{\rm MIR}$\,/\,L$_{\rm FIR}$ value than the
quasars (by a factor $\sim$3), the distribution of 
[NeV]$_{\rm 14.3 \mu m}$\,/\,[NeII]$_{\rm 12.8 \mu m}$ 
is strikingly similar for both samples:
  The logarithmically determined mean of the [NeV]\,/\,[NeII] ratios
  (and in bracketts the rms expressed as factor around this mean) are
  1.1 (2.8) and 0.9 (2.1) for the quasars and galaxies, respectively.
Furthermore, the 
[NeV]$_{\rm 14.3 \mu m}$\,/\,[NeII]$_{\rm 12.8 \mu m}$ 
values are much higher (by a factor $\sim$100) than those 
found for starburst galaxies 
(Sturm et al. 2002), placing also the FR\,2 galaxies clearly in the
AGN dominated range. We find also similar results using other lines, e.g.
[NeVI]$_{\rm 7.6 \mu m}$, [NeIII]$_{\rm 15.5 \mu m}$, [SIV]$_{\rm 10.5 \mu m}$.

If the unification between powerful galaxies and quasars 
is correct, then the ratio of central AGN luminosity to isotropic 
radio lobe power should be similar for both types.  
Fig.~\ref{fig3} (vertical axis) shows the luminosity ratio 
[NeV]$_{\rm 24.3 \mu m}$\,/\,P$_{\rm 178\,MHz}$. 
Again, the distributions look similar for both samples:  
  The logarithmically determined mean of the ratios
  (and in bracketts the rms expressed as factor around this mean) are
  0.027 (2.2) and 0.023 (4.1) for the quasars and galaxies, respectively.   
This suggests that the galaxies possess 
intrinsic AGN luminosities similar to those of the quasars. 
It provides direct spectroscopic evidence in favour of the unified
schemes.

How can these results be reconciled with the fact 
found by many authors, that galaxies show a
ten times lower  
[OIII]$_{\rm 500.7 nm}$\,/\,P$_{\rm 178\,MHz}$ luminosity ratio than 
quasars? 
Does optical extinction play a role 
as proposed by e.g. Jackson \& Browne (1990),
Hes et al. (1996), Baker (1997) and 
  di Serego Alighieri et al. (1997),
or do galaxies
possess a weaker central powerhouse suggested by e.g. Lawrence (1991) 
and Grimes et al. (2004)? 
We can now test these two possibilities using the [OIV]$_{\rm 25.9 \mu m}$
line:
it has a higher ionisation potential than the optical 
[OIII]$_{\rm 500.7 nm}$ line, but is $\sim$50 times 
less affected by extinction.
If the galaxies had intrinsically weaker AGN and hence weaker radiation 
fields, then
one expects their [OIII]$_{\rm 500.7 nm}$\,/\,[OIV]$_{\rm 25.9 \mu m}$ ratios
to be higher than for quasars. 
However, this is not the case,  rather we find the opposite as shown in 
Fig.~\ref{fig3} (horizontal axis). The galaxies exhibit a lower 
[OIII]$_{\rm 500.7 nm}$\,/\,[OIV]$_{\rm 25.9 \mu m}$ luminosity ratio
than all but one of the quasars.
  For this one quasar (3C\,298) the [OIII]$_{\rm 500.7 nm}$
  is highly uncertain, since extrapolated from the total
  (broad and narrow line) H$_{\alpha}$ luminosity (Tab.\,\ref{tab1}).
  The logarithmically determined mean of the [OIII]\,/\,[OIV] ratios
  (and in bracketts the rms expressed as factor around this mean) are
  16.9  (2.2) and 
   2.5  (2.2)
   for the quasars and galaxies, respectively, hence the distributions
   are different at more than 5$\sigma$.
Therefore we suggest that the   
galaxies suffer from substantial [OIII]$_{\rm 500.7 nm}$ extinction
(A$_{\rm V}$\,$\simgreat$\,3), 
at least in the central regions and that most of 
the - in principle extended  - NLR [OIII]$_{\rm 500.7 nm}$ 
luminosity originates from the central part. 
This result questions the former use of the
[OIII]$_{\rm 500.7 nm}$ emission as
  isotropic tracer for testing the unified schemes, 
in particular for establishing the receding torus model.
As proposed by Haas et al. (2004), it seems that  
not only the inner radius, but also the height of the torus
increases with growing luminosity.

In contrast to our high luminosity 3CR objects, 
for Seyfert galaxies the luminosity ratio
[OIII]$_{\rm 500.7 nm}$/FIR is the same for type\,1 and type\,2
when the objects are selected by L$_{\rm FIR}$ with a 25$\mu$m/60$\mu$m
colour criterion (Keel et al. 1994, Fig 3).
This suggests that the [OIII]$_{\rm 500.7 nm}$ emission is
isotropic for those objects of relatively low luminosity.  
On the other hand, the mostly radio-quiet Sloan AGN seem to indicate that the
ratio of type\,1s to type\,2s increases with luminosity, when selected by
[OIII]$_{\rm 500.7 nm}$ (Simpson 2005).  This
result can be qualitatively explained in the context of the unified
model without recourse to a luminosity-dependent torus opening angle if the
[OIII]$_{\rm 500.7 nm}$ emission is
anisotropic at high intrinsic optical/UV luminosity.   
For the radio galaxies, if the [OIII]$_{\rm 500.7 nm}$ emission
is mostly hidden, then relative to the broad
lines it should be almost as strong in polarised flux as in total flux.
  Notably in three of five narrow-line radio galaxies ($0.11<z<0.22$),
  di Serego Alighieri et al. (1997)
  found polarised [OIII]$_{\rm 500.7 nm}$ emission. 

One source (the anomalous object 3C\,459) has 
only rather high upper limits in the high-excitation [NeV] lines,
which are still consistent with the unification. 
Its continuum is probably 
mostly starlight from a post-starburst population (Miller 1981).
Also, though classified as an NLRG here, some authors claim to have detected
broad lines (see the "notes" in NED).
Notably, its IRS spectrum 
exhibits excited molecular hydrogen emission (H$_{\rm 2}$ S(1) at 17.03$\mu$m 
and H$_{\rm 2}$ S(3) at 9.67$\mu$m with
luminosities of 10$^{\rm 33.78}$ W and 10$^{\rm 34.34}$ W,
respectively) - to our knowledge 
the first such detection in a powerful FR\,2 radio galaxy.
Compared with [OIV]$_{\rm 25.9 \mu m}$ and 
P$_{\rm 178\,MHz}$ this source shows also an FIR luminosity 
excess (Haas et al. 2004). 
Since 3C\,459 is a compact steep spectrum radio source, 
we suggest that the excited molecular hydrogen emission originates from shocks 
due to the probably young expanding jet.

On the other hand, it is worth to note that the radio galaxy 3C\,295
with a very low L$_{\rm FIR}$\,/\,P$_{\rm 178\,MHz}$ (Meisenheimer et
al. 2001) lies still at the low end of the
[Ne\,V]\,/\,P$_{\rm 178\,MHz}$ distribution, but has a rather
intermediate [Ne\,V]\,/[Ne\,II] ratio; this indicates that despite the
AGN-typical line ratio 3C\,295 has exceptionally bright
radio lobes possibly caused by a special evolution or environment as
suggested by Haas et al. (2004). 

Clearly, 
our data set is small and the results should finally be 
corroborated by larger samples from the entire Spitzer IRS data base.
Nevertheless, since there are at most one ($\sim$15\%) or two outliers in the
distributions (Figures 2 \& 3),  the conclusions drawn from our 2$\times$7 objects
may be generalized to 85\% or at least 70\% of the entire 3CR sources,
i.e. the steep radio spectrum quasars and powerful galaxies with
10$^{\rm 26.5}$ W/Hz $\simless$ P$_{\rm 178\,MHz}$ $\simless$ 10$^{\rm 29.5}$ W/Hz
and $0.05 \simless z \simless 1.5$.
Also, 
the dispersion in the distributions is
still large and needs to be explored further, possibly 
allowing for recognizing
evolutionary trends among these complex sources. 

\acknowledgements

This work is based on observations with the Spitzer Space Telescope operated
by the JPL, Caltech, under
contract with NASA. This research was supported by the Nordrhein-Westf\"alische
Akademie der Wissenschaften. 
We thank the referee Robert Antonucci for his inspiring comments.

\end{document}